# Josephson junction array type $I$–$V$ characteristics of quench-condensed ultra thin films of Bi

G. Sambandamurthy*, K. DasGupta, V.H.S. Moorthy, N. Chandrasekhar

*Department of Physics, Indian Institute of Science, Bangalore 560 012, India*

## Abstract

In this communication we report studies of dc current–voltage ($I$–$V$) characteristics of ultra thin films of Bi, quench condensed on single crystal sapphire substrates at $T = 15$ K. The hysteretic $I$–$V$ characteristics are explained using a resistively and capacitively shunted junction (RCSJ) model of Josephson junction arrays. The Josephson coupling energy ($E_J$) and the charging energy ($E_c$) are calculated for different thickness ($d$) values. A low resistance state is found in the low current regime below the critical current, $I_c$. This resistance $R_0$ is found to have a minimum at a particular thickness ($d_c$) value. Reflection high energy electron diffraction (RHEED) studies are done on these films. A distinct appearance of a diffuse ring near $d_c$ is observed in the diffraction images, consistent with the recent STM studies (Ekinci and Valles, Phys. Rev. Lett., 82 (1999) 1518). These films show an irreversible annealing when temperature is increased. The annealing temperature ($T_a$) also has a maximum at the same thickness. Although the $R_s$ vs $T$ of quench-condensed Bi films suggest that the films are uniform, our results indicate that even in thick films, the order parameter is not fully developed over the complete area of the film. These results are discussed qualitatively. © 2000 Elsevier Science Ltd. All rights reserved.

## 1. Introduction

Two-dimensional (2D) electron systems have been extensively studied, both experimentally and theoretically, to understand the basic physical phenomena such as the onset of superconductivity in such systems, over last few decades [1–4]. Depositing the material onto a cold ($T \sim 4.2$ K) substrate, a technique pioneered by Buckel and Hilsch [5], has been used to investigate insulator–superconductor transitions as a function of thickness in Bi films. Superconductivity has been observed in such films in the temperature range 1–10 K [1,2]. In these films the microstructural disorder plays an important role in determining the transport properties of the system, through phenomena such as localisation.

These quench-condensed films have been modelled as arrays of Josephson junctions [6]. But the critical currents ($I_c$) and relevant energy parameters (Josephson coupling energy, $E_J$, charging energy, $E_c$ etc.) have not been determined as a function of thickness of films.

In this communication, we report experimental measurements of current–voltage ($I$–$V$) characteristics at $T < T_c$ of quench-condensed films of Bi and model them as resistively and capacitively shunted Josephson junction (RCSJ) arrays. The $I$–$V$ curves exhibit hysteresis. From these curves we calculate the parameters $E_c$, $E_J$ etc. as mentioned above. At low currents ($I < I_c$), contrary to intuition, we find a low resistance state. This non-zero resistance is measured at different thickness values and it is found to have a minimum at a particular critical thickness value ($d_c$).

The physical factors that control the transport at such low coverages strongly depend on how the film grows and the resulting structure. Recent STM studies [7,8] have thrown some light on the morphology of these quench-condensed films. We have carried out reflection high energy electron diffraction (RHEED) studies on the quench-condensed Bi films on highly oriented pyrolytic graphite (HOPG) sheets. The electron beam is incident on the substrate at a glancing angle (<few degrees) so that the momentum transfer from the high energy beam to the growing film is minimal, ensuring that the intrinsic film morphology is not perturbed by this probe. RHEED has an advantage that it averages over large areas in the film structure and give a global picture of the growing morphology. A distinct diffuse ring characteristic of amorphous/nanocrystalline material is observed in the films. At very low coverages ($d < d_c$) there is no feature in the diffraction image and near $d_c$ the amorphous ring starts to develop and intensifies as the film thickness is increased.

* Corresponding author.
  E-mail address: samband@physics.iisc.ernet.in (G. Sambandamurthy).

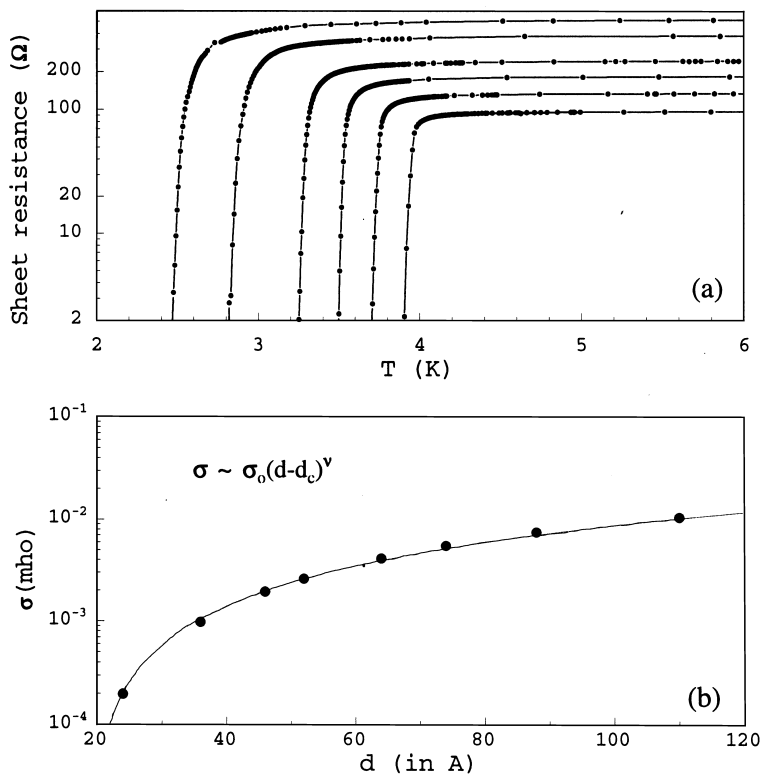

Fig. 1. (a) Evolution of sheet resistance $R_s$ vs $T$ for Bi films of different thicknesses. Films were quenched at $T_q = 15$ K. $T_c$ is mean field transition temperature. (b) Conductivity vs thickness of the films in (a). The conductivity follows the relation $\sigma \sim \sigma_0(d - d_0)^\nu$ with $\nu = 1.35$.

When the temperature of the films is raised, the films show an irreversible annealing of short range disorder at a characteristic temperature, depending on the thickness of the film. The resistance jumps to a higher value. This annealing temperature is measured for various thicknesses and it is found to have a maximum at $d_c$. We discuss these observations qualitatively.

## 2. Experimental details

The experiments were done in a UHV cryostat, custom designed for in situ experiments. The cryostat is pumped by a turbomolecular pump backed by an oil-free diaphragm pump. A completely hydrocarbon free vacuum $\leq 10^{-8}$ T can be attained. The substrate which is a single crystal (0001) sapphire of size $2.5 \times 2.5$ cm$^2$ is mounted on a copper cold finger whose temperature can be maintained down to 1.8 K by pumping on the liquid helium bath. The material (Bi) is evaporated from a Knudsen-cell with a pyrolytic boron nitride crucible, of the type used in molecular beam epitaxy (MBE). Bi is evaporated from the cell at 650°C. The material is evaporated into a four-probe resistivity measurement pattern by using a metal mask in front of the substrate. Successive liquid helium and liquid nitrogen cooled jackets surrounding the substrate reduce the heat load on the substrate and provide cryo-pumping. The metal flux reaching the substrate is controlled using a carefully aligned mechanical shutter in the nitrogen shield. The thickness of the film is increased by small amounts by opening the shutter for a time interval corresponding to the desired increase in thickness. A quartz crystal thickness monitor measures the nominal thickness of the film.

Electrical contacts to the film are provided through pre-deposited platinum contact pads (~50 Å thick). The wires leaving the cryostat are thermally anchored to the cold finger and to the walls of the cryostat. $I$–$V$'s and electrical resistance measurements are done using a standard dc current source and a nanovoltmeter.

When RHEED experiments were done, HOPG substrates were used since dielectric substrates like sapphire, quartz

## 3. Results and discussion

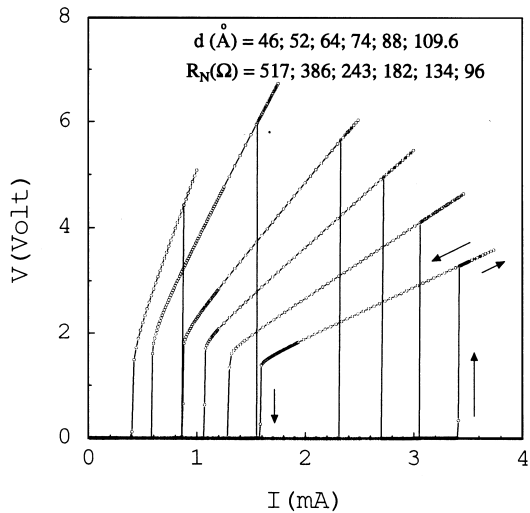

Fig. 2. *I–V* curves of the films, showing the hysteresis. The arrows indicate the direction of current sweep. The parameters obtained from these curves are listed in Table 1.

Fig. 1a shows the evolution of sheet resistance $R_s$ of the Bi films which were deposited on a sapphire substrate held at 15 K. The conductivity ($\sigma$) follows the relation $\sigma \sim \sigma_0(d - d_0)^\nu$ with $\nu = 1.35$ and $d_0 = 19$ Å (as shown in Fig. 1b), a value close to that of a 2D random resistor network near its percolation threshold [9]. The good fit to the percolation type conduction mechanism is an indication that the conduction in these films depends very much on the percolation path distributions, which are a consequence of the microstructure of the film. The current flow is percolative even after uniform coverage of the substrate, as discussed below. It is puzzling that this fit to percolation behaviour is observed for large thicknesses, as much as $5d_0$. This interesting observation will be addressed further in a separate publication.

The *I–V* characteristics were obtained at 2.25 K (which is below the $T_c$) for all the films. They are plotted in Fig. 2. When current is increased from the zero value, the voltage jumps to the normal state value at the critical current ($I_c$). Upon reducing the current from the normal state, the voltage returns to zero not at $I_c$, but at a much lower value $I_{min}$.

The above observations suggest that the film can be considered as an array of Josephson junctions, which are shunted by a resistance. Consequently the RCSJ model [10] can be used to describe the hysteretic behaviour of the *I–V* curves, with the capacitance being the intrinsic capacitance of the junction. From the ratio of $I_{min}/I_c$, the value of the admittance ratio ($\beta$) can be calculated get charged by the incident electron beam and results in beam deflection. This effect leads to artefacts in the diffraction pattern. We rationalise the use of HOPG substrates, since Ekinci and Valles [7,8] reported that the substrate does not significantly influence film morphology.

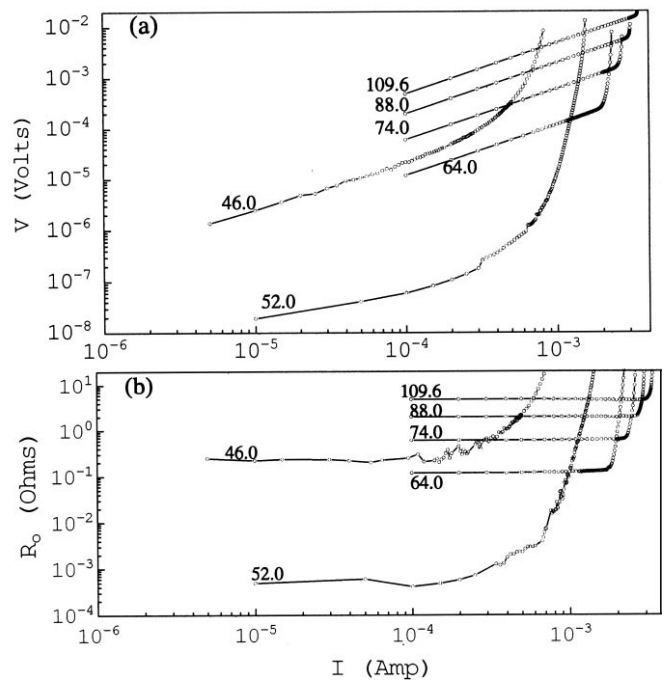

Fig. 3. (a) The same *I–V* curves in Fig. 2 in the low current ($I < I_c$) regime. Note this plot is on a log–log scale. (b) $R_0$ vs *I* for the different films. $R_0$ is calculated by finding the derivative of the *I–V* curves. $R_0$ shows a minimum at $d = 52$ Å.

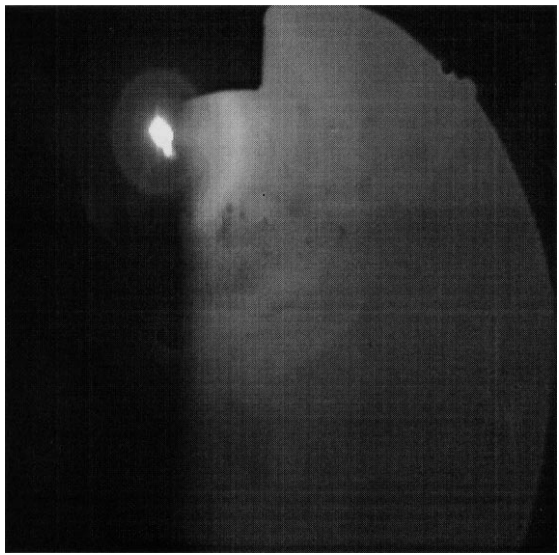

Fig. 4. The RHEED diffraction picture from a 82 Å thick Bi film on HOPG. At least one diffuse ring is clearly visible.

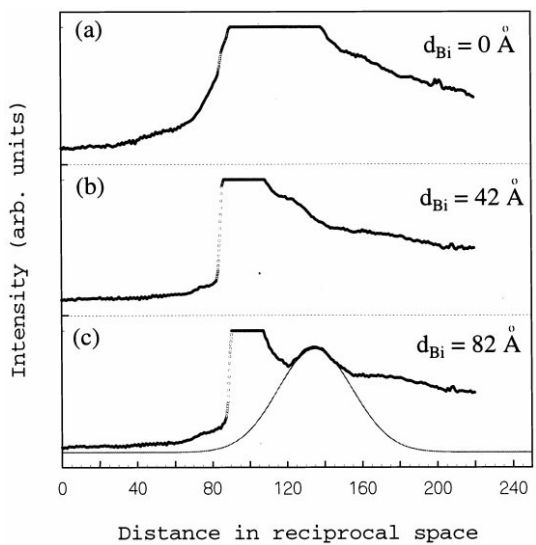

Fig. 5. The horizontal intensity profiles of the diffraction images with: (a) no Bi; (b) 42 Å Bi; and (c) 82 Å Bi on HOPG substrate. From the FWHM of the diffuse ring the grain size is calculated.

[11–13]. Here $\beta = \omega_c C/G$, where $\omega_c$ is $(2e/\hbar)I_c R_s$; $C$, the intergranular capacitance; and $G$, the normal state conductance of the array. We wish to point out that these are lumped parameters, which characterise the whole array. From the values of $\beta$, the intergrain capacitance is calculated. The charging energy $E_c(= e^2/2C)$ and the Josephson coupling energy $E_J(= \hbar I_c/2e)$ are calculated for all the film thicknesses studied. These values are listed in Table 1. These values are calculated using single value of $C$ and $G$, which correspond to capacitance and conductance of the array. $C$ and $G$ will have a range of values, the distribution of these values and the moments of the distribution will of course depend on the film thickness.

It is clear that these films are well into the classical underdamped junction regime with $E_J \gg E_c$ and the normal state sheet resistances $R_s$ much less than the quantum resistance, $R_Q = 6.45$ k$\Omega$. For all the films we find $E_J \gg k_B T$. (Under these conditions thermal fluctuation effects are unimportant [14], and both $I_c$ and $I_{min}$ are unaffected by thermal fluctuations.

The interesting observation in the $I$–$V$ curves is that for currents below $I_c$, the system is not in a zero-resistance state. The low current portion of the $I$–$V$ curves is shown in Fig. 3a. This resistance ($R_0$) is plotted as the function of the current in Fig. 3b.

This coexistence of superconductivity with resistance has been observed in 2D arrays of Josephson junctions [15] as well as in thin film systems [16]. This has been explained by the fact that there is always a finite probability for thermally generated quasi particles to tunnel across a junction, thereby disrupting the phase. This linear regime exists only for low currents $I \ll I_c$ for the thinnest films.

It is clear from the figure that at 52 Å thickness, $R_0$ has a minimum value and it increases as the thickness is increased. It means that at this particular value of thickness ($d_c$), the tunneling probability of the quasi-particles is a maximum for the thicknesses studied.

RHEED image from a 82 Å thick Bi film on HOPG is shown in Fig. 4. In Fig. 5 the horizontal intensity profiles of the diffraction images are plotted for bare HOPG substrate, 42 Å thick Bi film quenched at 15 K on HOPG and a 82 Å thick film. The diffuse ring structure clearly proves that the films are mostly amorphous in nature. From the full width at half maximum (FWHM) of the ring, the grain sizes are calculated to be around ∼75–100 Å. Recent STM studies [7,8,17,18] on the morphology of quench-condensed Pb films show that the film completes coverage of the substrate at ∼52 Å. Since these authors claim that the morphology is characteristic of the quench-condensation process, and is unaffected by the choice of deposition material (both Pb and Au) or substrate, we rationalise that around this thickness, our Bi films also cover the substrate completely. The persistence of the hysteresis in the $I$–$V$'s beyond this thickness indicates that despite the uniform coverage of the substrate, the current flow is percolative. This may be due to spatial non-uniformity of magnitude of the superconducting order parameter. Since the film consists of grains/clusters, there are a large number of grain-boundary junctions, and the superconducting order parameter has different values of phase in the grains across the junction.

These metastable films show an irreversible annealing to a stable crystalline structure when the temperature of the film is increased. Annealing studies were done on Bi films of different thicknesses. All the annealing studies were done on fire-polished quartz substrates. We have found that the annealing behaviour does not change whether the substrate is quartz or sapphire. A difference of $\pm 1$ K in the annealing temperature is found when the substrates were changed. Here we define the annealing temperature ($T_a$) as the temperature at which the sheet resistance of the film starts to increase sharply.

Table 1
Calculated values from RCSJ model

| $d$ (Å) | $T_c$ (K) | $I_c$ (mA) | $C$ ($\times 10^{-17}$ F) | $E_c$ ($\times 10^{-22}$ J) | $E_J$ ($\times 10^{-19}$ J) |
|---|---|---|---|---|---|
| 46 | 2.66 | 0.87 | 1.004 | 51.12 | 2.863 |
| 52 | 3.00 | 1.55 | 1.580 | 32.48 | 5.101 |
| 64 | 3.35 | 2.32 | 2.814 | 18.24 | 7.635 |
| 74 | 3.57 | 2.72 | 3.691 | 13.92 | 8.952 |
| 88 | 3.76 | 3.05 | 4.988 | 10.28 | 10.038 |
| 109.6 | 3.94 | 3.41 | 7.975 | 6.44 | 11.222 |

The annealing curves for different thickness films are presented in Fig. 6. The inset shows the variation of $T_a$ with the thickness of the film. Surprisingly for the film with thickness 53 Å, the annealing temperature is highest. Since Bi is a semi-metal, annealing to a higher resistance state is expected, since annealing of short range disorder would cause crystallisation, and deplete the density of states.

At $d_c$, the film seems to be in a more stable configuration than at other thickness values. A higher $T_a$ means that more thermal energy is needed to rearrange the atoms in the film to a stable configuration.

From the RHEED results and the earlier STM studies, we believe that around $d_c$, the film completes one layer of growth and starts to form the second layer. This means that when the first layer is complete, the resistance to thermally generated quasi-particle motion is minimum and more thermal energy is needed to disturb the stable configuration compared to the other thickness values. At this thickness, since the substrate coverage is complete, quasi-particle transport is easier, since at lower film thickness, the quasi-particles had to tunnel from one grain/cluster to another, which were separated by bare regions of the substrate which lowered the tunnelling probability. We wish to point out that vortices can exist in JJ-arrays, even in the absence of a magnetic field, since the discrete nature of the array can cause current circulation [19]. At this point it is not possible to determine whether the resistance that is observed is due to quasi-particles or vortices.

A JJ-array may also be composed of normal and superconducting regions, with the normal regions acting as weak links, in series with the superconducting regions. In this case too it would be possible to have a finite, temperature independent resistance at low temperatures due to quasi-particle tunnelling. The normal regions could be regions that have crystallised and become semi-metallic, like bulk Bi. Even if this explanation is entirely possible, we find no evidence for crystalline regions in our films, to the resolution of our RHEED.[1]

Although one may conclude that the quench-condensed Bi films are uniform or homogenous, by studying the $R_s$ vs $T$, this conclusion can be misleading. In uniform films suppression of amplitude of the superconducting order parameter is thought to be responsible for the vanishing of superconductivity in very thin films. Our results indicate that even in thick films, the order parameter is not fully developed over the complete area of the film. This spatial non-uniformity of the order parameter leads to the JJ-array type $I$–$V$ characteristics that we have reported. Thus these films are indeed granular, although the granularity does not show up in the $R_s$ vs $T$. Pb and Ga films are known to be granular from the $R_s$ vs $T$ itself, since a drop in the resistance occurs at the bulk $T_c$ even for thin films, which subsequently show insulating behaviour [20].

It must be borne in mind that the RCSJ model was developed for a single junction. We can, however, consider the $E_c$, $E_J$ values that we have evaluated, as lumped parameters for irregular arrays of the junctions that make up our films. It will be interesting to study the variation of this low value of $R_0$ at $d_c$ as function of temperature and magnetic field and compare the results with those obtained for fabricated junction arrays, in order to determine the origin of the resistance and dissipation. More experiments on the morphology of the film are also needed to understand structure–property correlations in these systems.

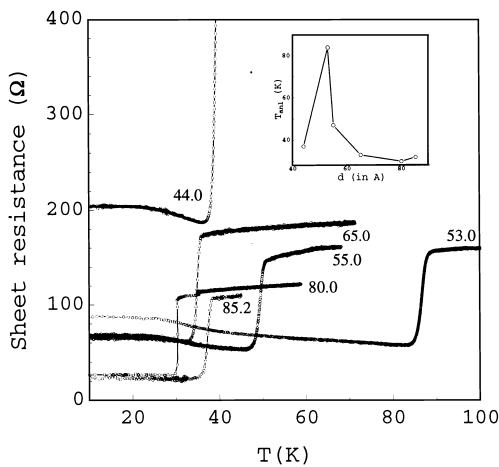

Fig. 6. Annealing behaviour of different films on quartz substrate. The 53 Å film anneals at a much higher temperature than the other films. Inset shows the dependence of the annealing temperature $T_a$ on the thickness of the films.

---

[1] We thank the referee for drawing our attention to this possibility.

**Acknowledgements**

This work is supported by the Department of Science and Technology, Government of India. We thank the referee for useful comments, and for drawing our attention to a possible alternative explanation of our results. One of the authors (KDG) thanks the Council of Scientific and Industrial Research, New Delhi for the fellowship.